\documentclass[journal=jacsat,manuscript=article]{achemso}
\usepackage{amssymb}
\usepackage{graphicx}
\usepackage{caption} 

\author{Hamdy Arkoub}
\affiliation{Department of Nuclear Engineering, The Pennsylvania State University, University Park, PA, 16802, USA}
\author{Jia-Hong Ke}
\affiliation{Computational Mechanics and Materials Department, Idaho National Laboratory, Idaho Falls, ID, 83415, USA}
\author{Miaomiao Jin}
\email{mmjin@psu.edu}
\affiliation{Department of Nuclear Engineering, The Pennsylvania State University, University Park, PA, 16802, USA}

\title{Atomistic Mechanisms of Stress‑Dependent Molten Salt Corrosion in NiCr Alloys}

\begin{document}

\begin{center}
\textbf{Table of Contents (TOC) Graphic}
\includegraphics[width=0.8\textwidth]{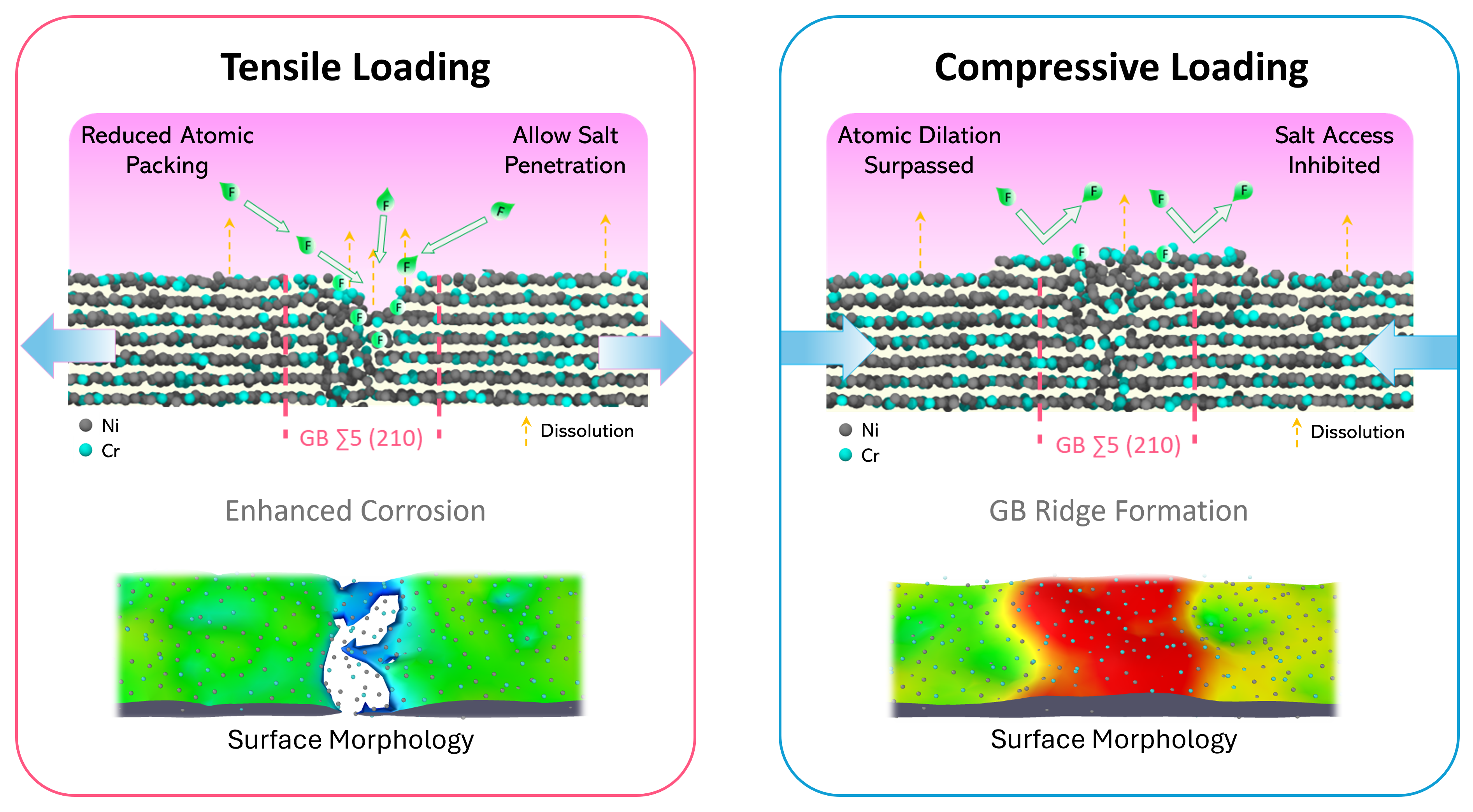}
\end{center}

\newpage
\begin{abstract}
Ni-based structural alloys in molten salt environments often experience simultaneous mechanical loading and corrosive attack, yet the mechanisms governing stress\-corrosion interactions remain unclear. Prior studies largely emphasize tensile stress, while the role of compressive stress has received limited attention. Here, reactive molecular dynamics simulations are used to investigate the coupled effects of applied strain and corrosion in Ni$_{0.75}$Cr$_{0.25}$ exposed to molten FLiNaK at 800$^\circ$C. A $\Sigma5(210)$ grain boundary model is subjected to tensile (+4\%) to compressive (-4\%) uniaxial strains, and corrosion behavior is evaluated through fluorine adsorption, charge redistribution, and grain boundary evolution. Tensile strain accelerates intergranular corrosion by reducing local atomic packing through elastic dilation and increasing excess free volume at the grain boundary, which enhances atomic mobility and salt infiltration. In contrast, compressive strain suppresses corrosion by promoting the formation of a ridge-like surface layer along the grain boundary, limiting salt access to the underlying alloy. These results provide atomistic insight into how stress states influence grain boundary corrosion in molten salts.

\end{abstract}


\section{Introduction}

In advanced energy systems such as molten salt reactors or concentrated solar energy storage, Ni-based structural alloys must withstand high temperatures, aggressive molten halide chemistry, and sustained mechanical loading, often under complex multiaxial stress states \cite{yvon2009structural,kelly2014generation,serp2014molten,xu2025molten}. These stresses can arise from various sources, such as thermal gradients, internal pressure loading, and fabrication-induced residual stresses \cite{busby2019technical,aitkaliyeva2017irradiation,bell2019corrosion,amiri2025understanding}. Molten fluoride salts, such as FLiNaK, are attractive coolants and heat-transfer media because of their high thermal stability and heat capacity \cite{williams2006assessment,delpech2010molten}. However, protective oxides are unstable in these environments, enabling fluorine-driven selective dissolution of Cr (e.g., CrF$_2$/CrF$_3$ formation), which leads to dealloying and progressive degradation of Ni-based alloys \cite{sohal2010engineering,olson2009materials,chan2022insights,arkoub2025first,arkoub2025surface}. 

Although molten-salt corrosion has been widely studied, its coupling with mechanical stress remains poorly understood \cite{xu2025molten,yvon2009structural}. While tensile stress is well recognized to promote intergranular cracking in aqueous environments \cite{amiri2025understanding}, experimental studies of stress corrosion cracking (SCC) in molten salts are limited by the challenges of applying controlled loads at high temperature in aggressive halide melts. Existing molten-salt SCC studies focus primarily on tensile loading and report accelerated Cr dissolution, enhanced intergranular penetration, and increased corrosion rates under tension \cite{fu2015impacts,gu2022stress,liang2025understanding}. However, these studies rely mainly on macroscopic metrics such as mass loss and crack formation and do not resolve how stress influences corrosion kinetics and grain boundary stability, a limitation also recognized in the recent review by Xu et al. \cite{xu2025molten}. This knowledge gap is important because stress fields in practical applications are highly heterogeneous. Thermal gradients, residual stress, high temperature creep, and operational transients generate both tensile and compressive regions \cite{amiri2025understanding}. Recent four-point bending experiments in FLiNaK suggest that compressive stress can reduce corrosion susceptibility in stainless steels relative to tensile regions \cite{williams2025combining}, yet the origin of this apparent protective effect remains unclear. Similar trends have been reported in aqueous SCC systems, where compressive stress can suppress crack initiation and slow crack growth by affecting near-surface transport and interfacial chemistry \cite{jayaraman2005comparison,ming2022improving,chu2023mechanism,xiao2025insights}. Whether comparable mechanisms operate in molten salts remains an open question.

Direct experimental observation of early-stage corrosion in molten salts, particularly under stress, is extremely challenging. Density functional theory (DFT) calculations can provide insights into initial adsorption and reaction energetics, but their computational cost limits access to dynamic corrosion processes \cite{yin2018first,startt2021ab,schneider2024mechanistic}. Reactive molecular dynamics (RMD) with ReaxFF enables chemically reactive, large-scale simulations that capture bond formation and dissolution at elevated temperatures \cite{van2001reaxff, senftle2016reaxff}. Prior RMD studies have shown that corrosion of Ni–Cr alloys in molten FLiNaK is driven by strong Cr–F bonding, leading to preferential Cr dissolution, and that the process is primarily controlled by near-surface atomic transport in bulk systems \cite{arkoub2024reactive,arkoub2025surface}.   

In this work, RMD simulations are used to investigate how applied strain/stress affects corrosion of a Ni$_{0.75}$Cr$_{0.25}$ alloy in contact with FLiNaK. A $\Sigma 5$(210) grain boundary model is subjected to uniaxial tensile (+4\%) and compressive (-4\%) strains. Interfacial reactivity and atomic mobility are quantified to determine how strain state influences early-stage intergranular corrosion. By comparing various strain states under otherwise identical conditions, we isolate the fundamental role of stress in governing early-stage corrosion mechanisms.

\section{Methods}

RMD simulations were performed using LAMMPS \cite{thompson2022lammps}, using a previously developed Reax\-FF force field for the NiCr–FLiNaK system \cite{arkoub2024reactive}. This potential reproduces key thermophysical properties of FLiNaK and captures preferential Cr dissolution in Ni–Cr alloys exposed to molten fluorides \cite{arkoub2024reactive,arkoub2025surface}. A $\Sigma5(210)/(001)$ symmetrical tilt GB in FCC Ni, with 53.1° tilt angle, was constructed using the coincidence site lattice (CSL) approach \cite{watanabe1984approach} and generated with the Aimsgb Python package \cite{cheng2018aimsgb}. The Ni$_{0.75}$Cr$_{0.25}$ alloy is created by randomly substituting Ni atoms with Cr to achieve 25\% Cr in each layer. The bicrystal model contains 18 layers along the $z$-direction and was first relaxed under NPT at 800$^\circ$C and 1 atm for 50 ps with a 0.25 fs time step using the Nose–Hoover thermostat and Berendsen barostat\cite{berendsen1984molecular,nose1984unified}. 

\begin{figure}[!ht]
	\centering
	\includegraphics[width=1.0\textwidth]{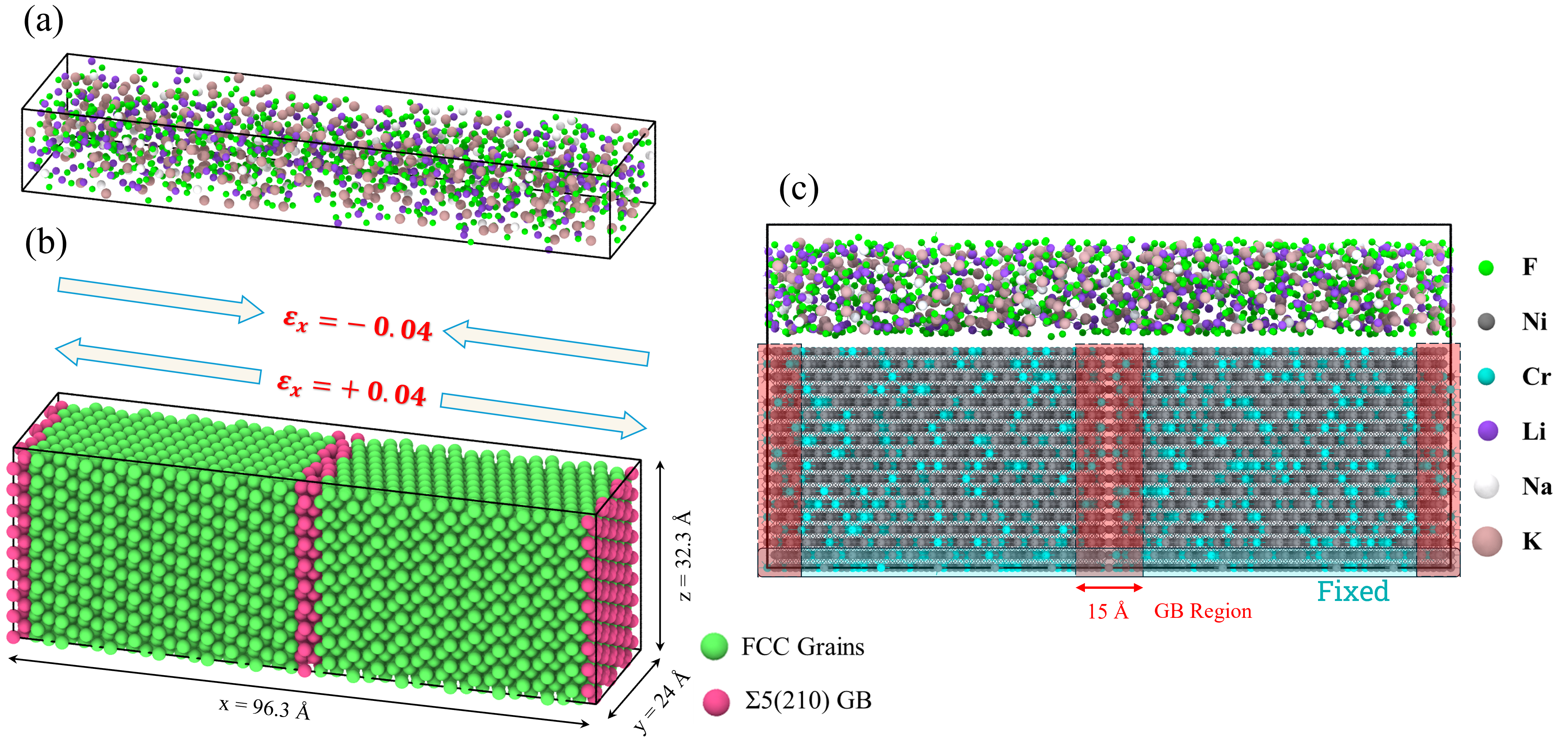}
	\caption{Atomic configurations of the optimized (a) FLiNaK salt and (b) Ni${_{0.75}}$Cr${_{0.25}}$ $\Sigma5(210)$ GB slab used in the study, respectively. (c) shows the alloy-salt model used in MD simulations, where the red-shaded zone denotes the GB region, defined as a 15 {\AA} wide slab centered on the GB plane.}
	\label{fig:model}
\end{figure}

As shown in Figure \ref{fig:model}, to generate strained configurations, the alloy slab was deformed along the $x$-direction at a constant strain rate of 10$^{-5}$ ps$^{-1}$ at 800$^\circ$C. Periodic boundary conditions were applied in x and y, and non-periodic conditions in $z$. Loading simulations were performed in the NVT ensemble. The stress–strain response (Figure S1 in Supporting Information (SI)) is linear over the simulated strain range, similar to other metal systems in MD simulations \cite{huang2023atomistic,ma2023applied,li2020molecular}. Three models in the elastic range were selected for corrosion simulations: tensile strain (+4\%), compressive strain (-4\%), and unstrained (0\%). The selected strain magnitude is larger than typical macroscopic service strains, but can reveal stress-dependent differences within the limited temporal scales accessible to MD simulations. Because the response remains elastic, the imposed deformation represents a well-defined tensile or compressive stress state without introducing plastic deformation. The corresponding cell sizes and stresses are listed in SI Table S1. Molten FLiNaK was constructed using Packmol \cite{martinez2009packmol} with a composition of 46.2–11.5–42 mol\% LiF–KF–NaF \cite{williams2006assessment}, corresponding to 363 LiF, 90 KF, and 328 NaF molecules. The salt was equilibrated separately at 800$^\circ$C and 1 atm with a 0.25 fs timestep, while constraining the $x$ and $y$ dimensions to match the alloy slab. The equilibrated salt was then placed on the alloy surface to form the alloy–salt system (Figure \ref{fig:model}c). The metal slab was partitioned into two regions: a GB region defined as a 15~\AA{} slab centered on the GB plane, and the remaining interior region referred to as bulk.

Corrosion simulations were performed at 800$^\circ$C using the Nose–Hoover thermostat \cite{nose1984unified}. This temperature is representative of molten salt reactor operating conditions and enables observable corrosion evolution within RMD timescales \cite{wright2018status,serp2014molten,arkoub2025surface}. To better approximate bulk constraint, the bottom two alloy layers were fixed. Fluoride surface coverage was quantified using cutoff distances set to 80\% of the sum of van der Waals radii: 2.776~\AA{} for Cr--F and 2.48~\AA{} for Ni--F, consistent with prior work \cite{arkoub2024reactive,arkoub2025surface}. A metal atom was classified as dissolved when it had fewer than two neighboring metal atoms and formed stable metal-F bonds. Ten independent simulations were performed for each strain state to ensure statistical reliability. Visualization and post-processing were carried out using OVITO \cite{stukowski2010visualization}.

\section{Results and Discussion}

\begin{figure}[!ht]
	\centering
	\includegraphics[width=1.0\textwidth]{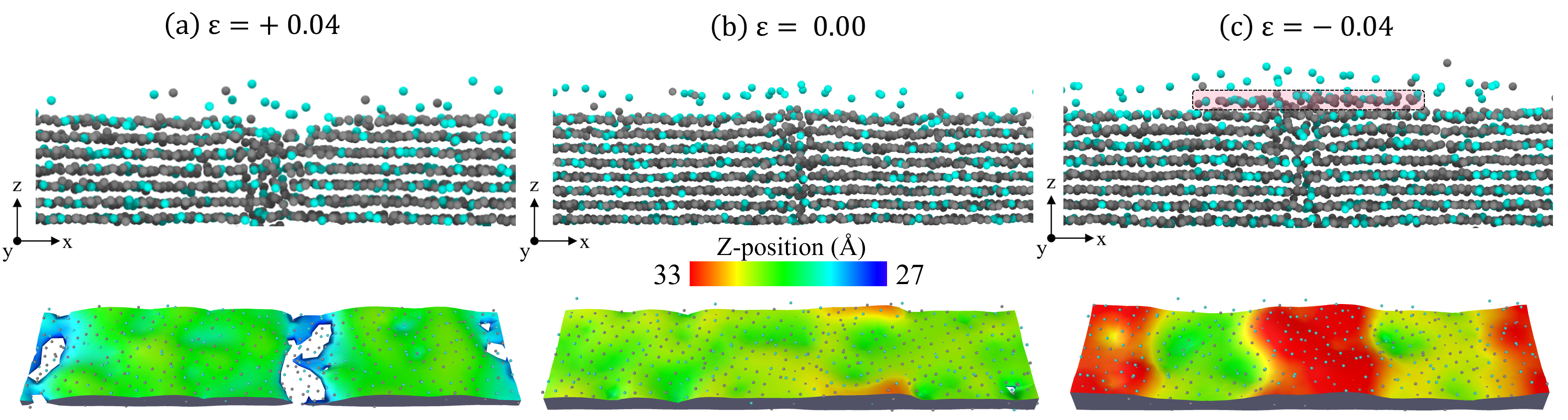}
	\caption{Zoomed-in atomic snapshots (top) and surface meshes (bottom) of the Ni${_{0.75}}$Cr${_{0.25}}$ $\Sigma5(210)$ GB slabs after 500 ps of exposure to molten FLiNaK under (a) tensile strain (+0.04), (b) zero strain, and (c) compressive strain (–0.04). Surface meshes are colored by z-position, while Ni and Cr atoms are shown in gray and cyan, respectively. }
	\label{fig:surfaces}
\end{figure}

Figure \ref{fig:surfaces} compares atomic configurations and surface morphology after corrosion under the three strain states. The surfaces highlight changes associated with early-stage stress-assisted corrosion at the GB. Under tensile strain (Figure \ref{fig:surfaces}a), pronounced localized penetration along the GB is observed, consistent with stress-accelerated intergranular attack as observed in Gu et al.\cite{gu2022stress}. In contrast, under compressive strain (Figure \ref{fig:surfaces}c), a localized protrusion develops at the GB, forming a ridge-like feature. The unstrained case (Figure \ref{fig:surfaces}b) shows a relatively uniform surface with only minor roughening and no pronounced GB recession or protrusion. Supporting Figure S2 shows that the total number of dissolved atoms over 500 ps is similar for all strain states. This indicates that the current timescale captures early-stage redistribution of corrosion rather than large differences in total mass loss. However, in all cases, the number of dissolved atoms originating from the GB region is nearly twice that from the bulk region, confirming preferential intergranular corrosion. This preferential GB dissolution increases local free volume and may influence subsequent mechanical stability.

It is worth noting that this GB ridge formation under the compressive condition is absent in the strained alloy equilibrated at 800$^\circ$C prior to salt exposure (Figure S1) and starts to develop within approximately 100 ps after contact with the melt. Hence, it arises from coupling corrosion-induced lattice disruption and compressive loading, rather than purely mechanical deformation. In the corrosion stage, fluorine adsorption and CrF$_x$ formation are active, leading to Cr detachment in the form of CrF$_x$ species into the melt. Selective Cr removal generates excess free volume along the GB, which enhances diffusion, while compressive stress promotes atomic flux toward the free surface. The combined effect produces outward mass transport and accumulation of material along the GB trace, forming a secondary ridge layer. Similar compressive-driven GB ridge formation has been reported in thin metal films, where stress is accommodated by surface extrusion \cite{genin1995effect,wang2023emergent}.

\begin{figure}[!ht]
	\centering
	\includegraphics[width=1.0\textwidth]{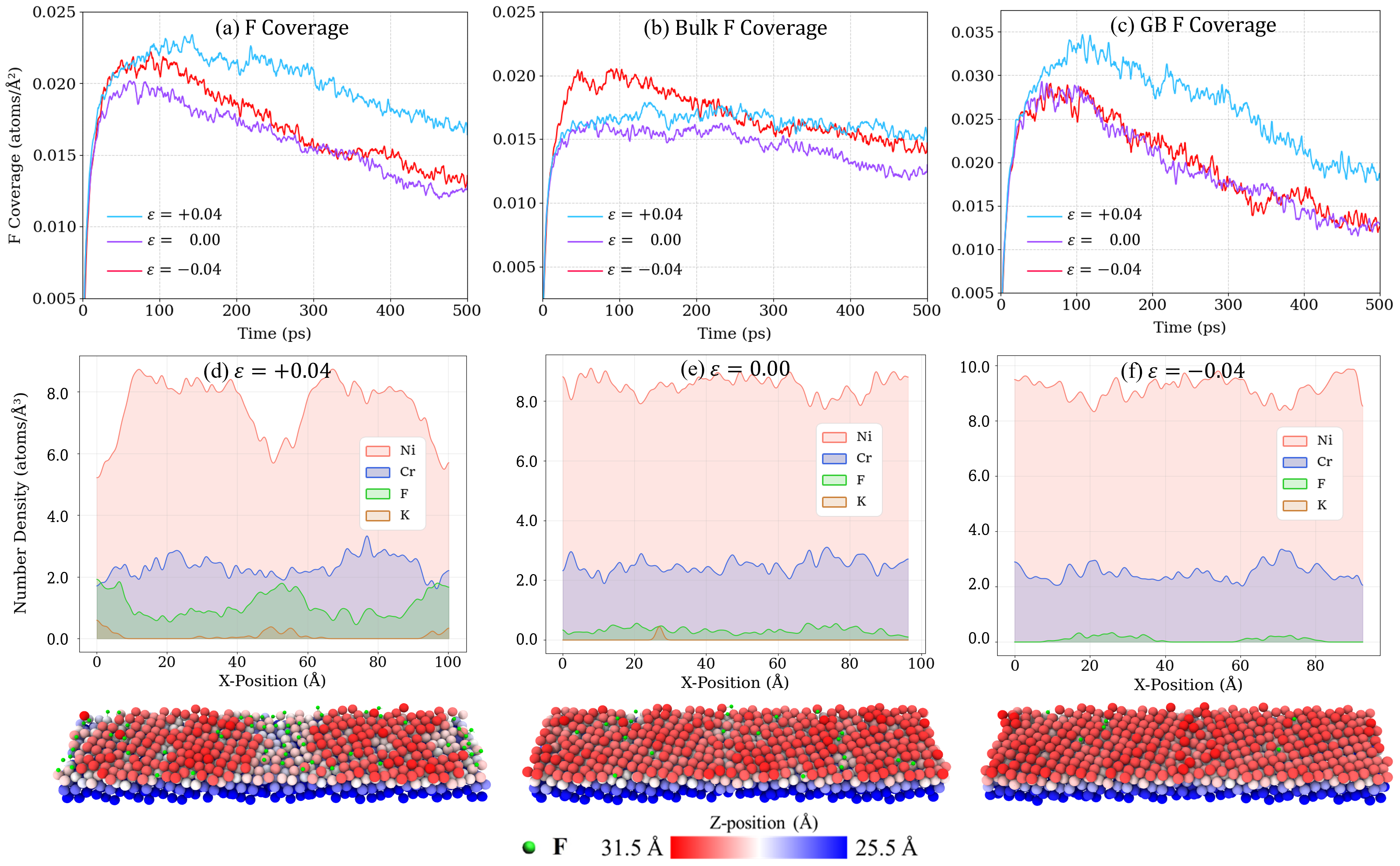}
	\caption{(a–c) Time evolution of F surface coverage averaged over the entire surface, bulk regions, and GB regions, respectively, for tensile, unstrained, and compressive cases. (d–f) Lateral number-density profiles at 500 ps along the x-direction for atoms within the near-surface slice defined by $Z$ = 25.5-31.5 {\AA} for tensile, unstrained, and compressive conditions, respectively, while bottom row show corresponding atomistic snapshots of the same near-surface region, with F atoms highlighted and atoms colored by z-position.}
	\label{fig:coverage}
\end{figure}

Since fluorine–metal interaction governs the initial corrosion reaction, Figure \ref{fig:coverage} quantifies the evolution of fluorine surface coverage under the three strain states. When averaged over the entire surface (Figure \ref{fig:coverage}a), the tensile case exhibits the highest F coverage, whereas the unstrained and compressive cases remain lower. This indicates that tensile strain increases the availability or reactivity of adsorption sites. The difference is primarily localized at the GB. Figure \ref{fig:coverage}c shows significantly higher fluorine coverage in the GB region under tension compared with the other two cases, while the bulk region (Figure \ref{fig:coverage}b) displays only minor variation among strain states. To resolve the spatial distribution, Figures \ref{fig:coverage}d–f present fluorine number-density profiles along the x-direction within the near-surface region ($Z$ = 25.5-31.5 {\AA}) at 500 ps, together with corresponding atomic snapshots. Under tensile strain (Figure \ref{fig:coverage}d), fluorine exhibits a pronounced density peak at the GB location. This localization is consistent with enhanced adsorption at dilated boundary sites. This GB-localized fluorine ingress under tension explains the morphology observed in Figure \ref{fig:surfaces}a. Tensile loading dilates the boundary and increases GB free volume. The Ni density profile shows a clear dip at the GB, reflecting reduced atomic packing associated with boundary opening under tension. These changes create energetically favorable sites for fluorine adsorption and ingress along the boundary. The resulting F accumulation further drives selective Cr dissolution, weakens the GB cohesion, and promotes intergranular corrosion. This interpretation is consistent with the localized recession observed under tension and aligns with experimental reports that tensile stress increases GB corrosion susceptibility in molten FLiNaK \cite{gu2022stress,williams2025combining}. 

The unstrained case (Figure \ref{fig:coverage}e) shows a relatively uniform lateral population of fluorine within the surface region, without a distinct GB peak at this stage, indicating that fluorine ingress is not strongly localized. Under compression (Figure \ref{fig:coverage}f), the surface F population is markedly altered. The snapshot shows reduced fluorine accumulation at the GB, and the x-profile shows little to no fluorine enrichment at the boundary. This attenuation of local F coverage coincides with the GB ridge layer formed under compression (Figure \ref{fig:surfaces}c). The protruded layer acts as a barrier that partially limits salt penetration to the GB and reduces localized Cr-F bonding. As a result, intergranular corrosion is comparatively suppressed under compressive strain.

\begin{figure}[!ht]
	\centering
	\includegraphics[width=1.0\textwidth]{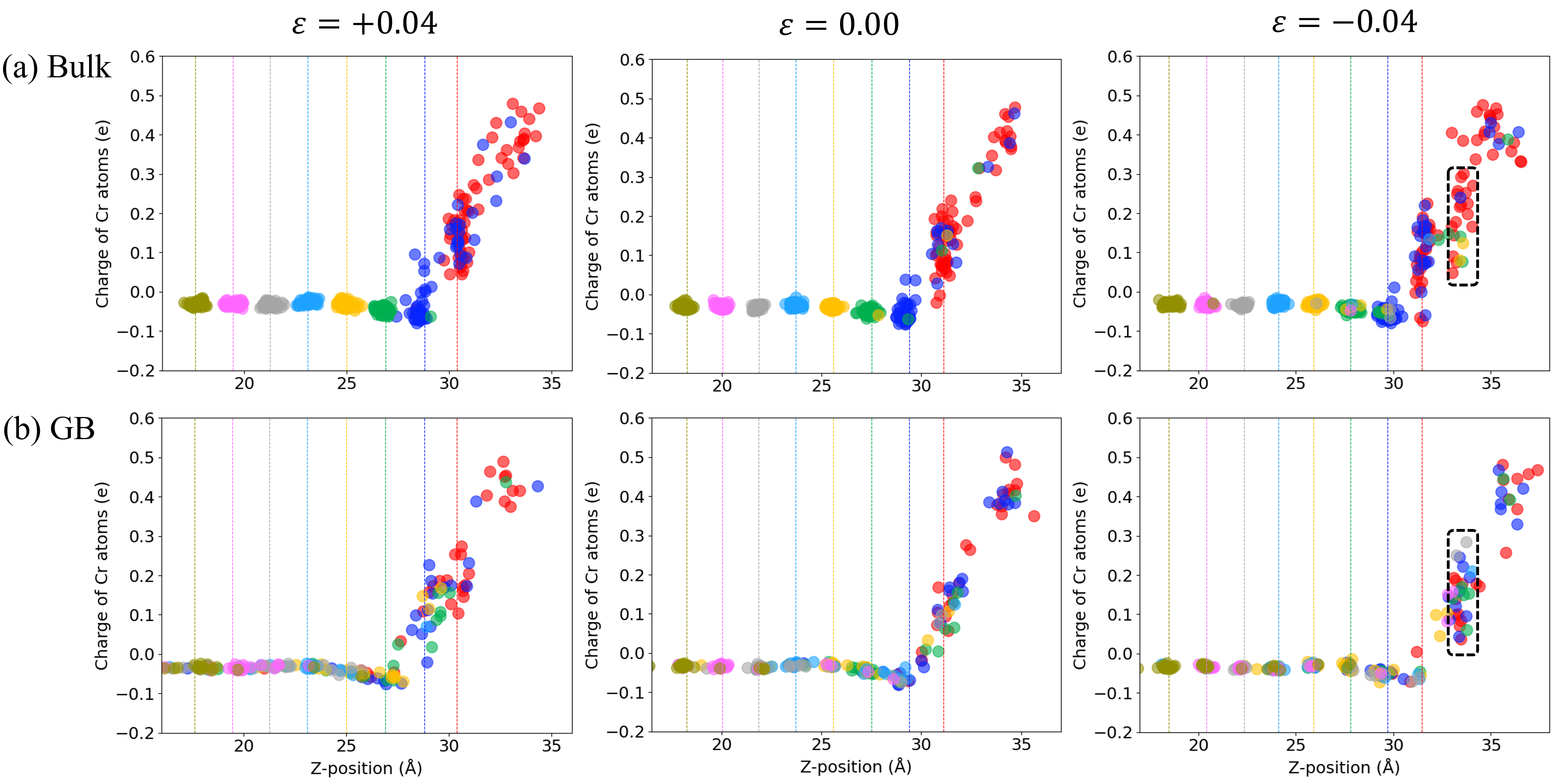}
	\caption{Distribution of Cr atomic charges along the z-direction after 500 ps of exposure to molten FLiNaK at 800 °C  for the three strain states. (a) Top panels show bulk regions, while (b) bottom panels correspond to GB regions. Points are colored according to each atom’s original layer in the pristine structure, and vertical dashed lines mark the initial layer positions.}
	\label{fig:charges_Cr}
\end{figure}

Figure \ref{fig:charges_Cr} compares Cr charge distributions in the bulk and GB regions under the three strain conditions after corrosion. Atoms are color-coded by their original layer in the pristine structure, allowing tracking of interlayer migration. In all cases, deeper-layer metal atoms retain near-zero charge, while significant positive charges develop near the salt–metal interface due to Ni/Cr–F interactions, consistent with prior ReaxFF studies of NiCr–FLiNaK systems \cite{arkoub2024reactive,arkoub2025surface}. A clear difference emerges between bulk and GB regions. In the bulk, Cr atoms largely remain confined to their original layers, and charge perturbations are restricted to the topmost surface. This behavior indicates that corrosion remains surface-limited away from the boundary. At the GB, however, substantial interlayer mixing is observed. Cr atoms originating several layers below the surface migrate upward and acquire a positive charge as they bond with fluorine. This vertical redistribution indicates that the GB acts as a fast transport pathway, supplying Cr from the interior to the reacting interface.

The extent of the GB transport depends on the strain state. Under tensile strain, GB atomic positions become broadly distributed along the $z$-direction, and positively charged Cr atoms appear deeper into the structure due to enhanced fluorine infiltration along GB. In contrast, under compressive strain, interlayer migration remains active but is more spatially confined. The secondary ridge layer contains GB atoms originating from multiple layers due to outward mass transport toward the surface. However, positively charged Cr atoms are largely concentrated near this reconstructed region instead of extending deeply along the boundary. Hence, compressive loading slows down salt penetration along the GB plane, reducing intergranular corrosion. Interestingly, while GB penetration is mitigated under compression, surface disruption away from the boundary appears more pronounced in the bulk region. This suggests that compressive stress redistributes corrosion activity from localized boundary attack toward more distributed surface reactions. Ni atomic charge distributions exhibit the same strain-dependent transport trends observed for Cr as visualized in the SI Figure S3.

\begin{figure}[!ht]
	\centering
	\includegraphics[width=0.8\textwidth]{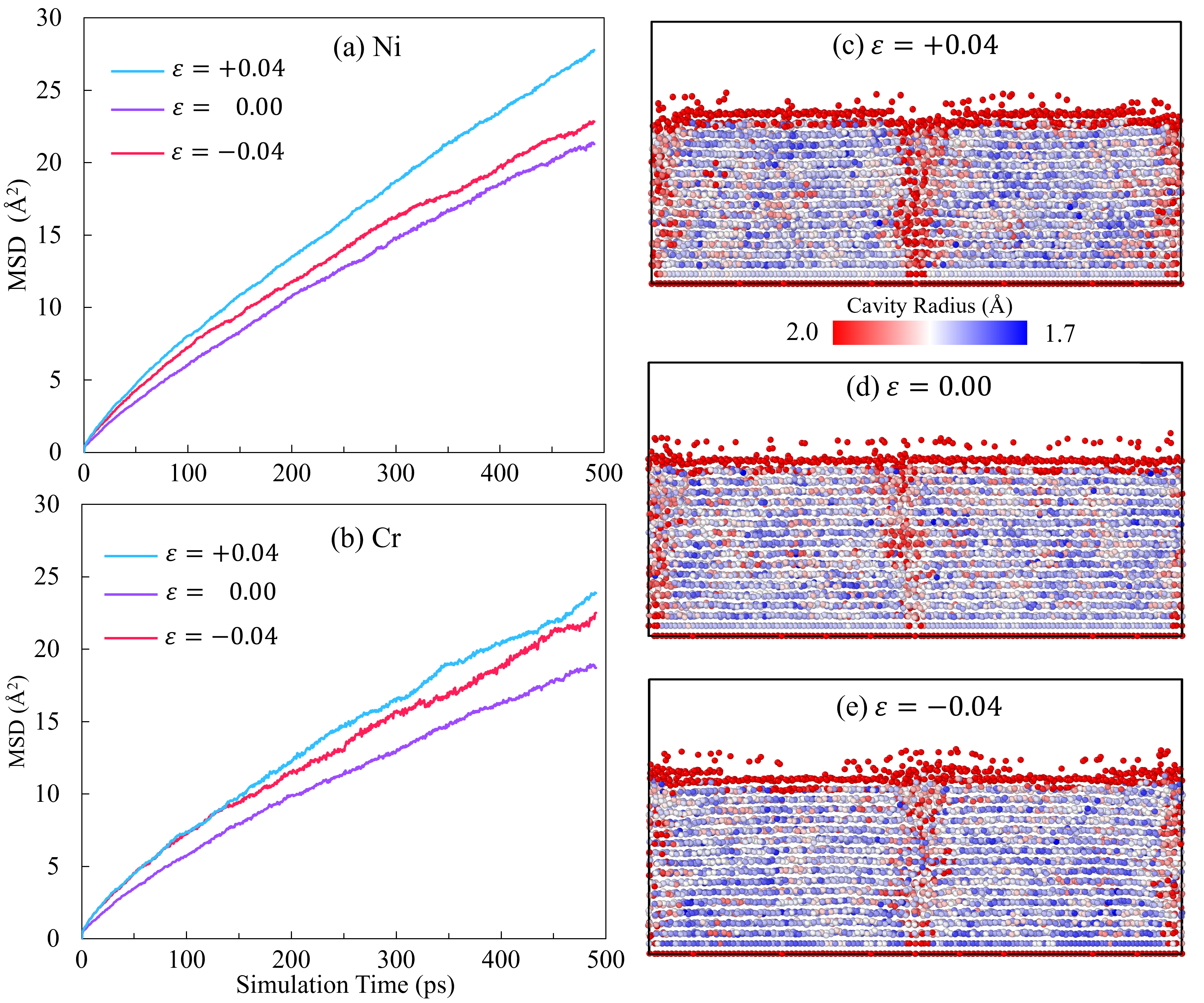}
	\caption{(a-b) MSD of GB Ni and Cr atoms as a function of time at 800$^\circ$C for the three strain states, respectively. (c-e) Final atomic configurations after 500 ps of corrosion, colored by Voronoi cavity radius to represent local free volume, for tensile (+4\%), unstrained (0\%), and compressive (-4\%) conditions, respectively.}
	\label{fig:msd}
\end{figure}

To quantify atomic mobility, the mean square displacement (MSD) of GB-region Ni and Cr atoms was evaluated at 800$^\circ$C (Figures \ref{fig:msd}a–b), and diffusion coefficients were extracted from the linear MSD regime (details in SI). The corresponding diffusion coefficients are listed in Table S1. Both species exhibit the highest mobility under tensile strain, intermediate mobility under compression, and the lowest values in the unstrained case. The enhancement under tensile strain is consistent with boundary dilation and increased free volume, which reduce migration barriers and promote interlayer transport. More interestingly, compression also leads to higher GB mobility than the unstrained case. This behavior indicates that compressive loading does not reduce GB mass transport. Under compression, selective Cr dissolution generates excess free volume at the GB, and stored elastic strain energy provides a driving force for atomic rearrangement. The resulting mobility is largely directed toward vertical mass redistribution and ridge formation. Notably, the GB diffusion coefficients obtained here exceed previously reported near-surface diffusion values for NiCr in FLiNaK at 800$^\circ$C \cite{arkoub2025percolating}. This suggests that the GB acts as an even more efficient mass-transport pathway than lateral surface diffusion. Such efficient GB transport is consistent with the experimentally observed tendency for intergranular corrosion in NiCr alloys \cite{bawane2022visualizing,yang2023one}.

Related to the strain-dependent transport behavior discussed above, prior experimental studies have shown that tensile stress can enhance Cr and Ni diffusivities in steels and Ni-based alloys by modifying chemical potentials and migration barriers in stressed lattices \cite{liu2022effects,gu2022stress,fu2015impacts,douglass1996international,rahmel1985international}. In Fu’s molten salt corrosion study of austenitic steel, tensile stress was suggested to promote defect formation and accelerate Cr transport to the surface for oxide growth \cite{fu2015impacts}. In this work, there was no dislocation emission or lattice defect generation during the corrosion, that contributes to diffusion. The enhanced transport under tension originates from elastic lattice dilation and increased excess free volume at the GB rather than defect-assisted diffusion. Similar understanding of stress-modified diffusion behavior without invoking bulk defect formation was noted in phase-field simulations of oxygen transport in titanium \cite{wang2024investigation}.

The structural differences under different strain states are shown in Figure \ref{fig:msd}c–e based on Voronoi free-volume analysis \cite{stukowski2010visualization,du1999centroidal}. In all cases, the $\Sigma5$ GB exhibits higher free volume than the grain interior, as expected for a high-angle boundary. Under tensile strain (Figure \ref{fig:msd}c), the GB displays significantly larger cavity radii and a broader region of elevated free volume compared with the unstrained and compressive cases. This dilation increases the availability of low-coordination sites and reduces migration barriers, consistent with the enhanced GB diffusion and fluorine localization observed earlier. Similar tension-driven vacancy and free-volume enrichment at GBs has been linked to SCC in atomistic simulations of $\alpha$-Fe in supercritical water\cite{huang2023atomistic}. In contrast, the unstrained (Figure \ref{fig:msd}d) and compressive (Figure \ref{fig:msd}e) cases exhibit comparable intrinsic GB free volume, with smaller cavity radii than the tensile case. Notably, compressive loading does not significantly reduce boundary excess free volume relative to the unstrained state. This is because compression is partially accommodated by outward mass transport and ridge formation at the surface, rather than by uniform densification of the GB core.  

\section{Conclusion}

Using reactive molecular dynamics simulations, this study shows how applied strain governs early-stage stress corrosion mechanisms in NiCr alloy exposed to molten FLiNaK at 800$^\circ$C. For a representative high-energy $\Sigma5(210)$ GB, tensile strain induces boundary dilation and increases excess free volume at the GB. This microstructural change enhances fluorine ingress and promotes mass transport along the GB plane, which accelerates Cr supply from subsurface layers to the salt–metal interface and hence accelerates intergranular penetration. In contrast, under compressive strain, stress is accommodated through outward mass transport and ridge formation along the GB. This surface reconstruction reduces direct salt access along the boundary plane and suppresses fluorine localization at the GB, thereby slowing intergranular attack. As a result, corrosion under compression shifts toward more spatially distributed surface degradation. These atomistic insights clarify stress–corrosion coupling in molten fluoride salts and highlight the importance of local stress state in determining alloy degradation pathways in molten salt environments.
 
\section*{Acknowledgments}
This research was funded by the National Science Foundation (NSF) through CAREER Award No. 2340019. The views, findings, conclusions, or recommendations presented in this work are solely those of the authors and do not necessarily represent those of the NSF. One of the authors (J.H.K.) would like to thank the support from the Laboratory Directed Research and Development (LDRD) Program at Idaho National Laboratory.

\section*{Data Availability}
This study did not involve the generation or analysis of datasets. Accordingly, no data are available to share.

\bibliography{references}



\end{document}